\documentclass[aps,prl,twocolumn,groupedaddress]{revtex4}
\usepackage{graphicx}
\usepackage{amssymb}

\begin{document}

\title{Critical exponent for the quantum Hall transition}

\author{Keith Slevin}
\affiliation{Department of Physics, Graduate School of Science, 
Osaka University, 1-1 Machikaneyama, Toyonaka, Osaka 560-0043, Japan}

\author{Tomi Ohtsuki}
\affiliation{Department of Physics, 
Sophia University, Kioi-cho 7-1, Chiyoda-ku, Tokyo 102-8554, Japan}

\date{\today}

\begin{abstract}
We report an estimate $\nu  = 2.593$ $\left[ {2.587,2.598} \right]$ of the critical 
exponent of the Chalker-Coddington model of the integer quantum Hall effect that 
is significantly larger than previous
numerical estimates and in disagreement with experiment.
This suggests that models of non-interacting electrons cannot explain the critical 
phenomena of the integer quantum Hall effect. \end{abstract}


\maketitle

The most important characteristics of a continuous phase transition are its 
critical exponents.  The critical exponents are a quantitative characteristic of 
the critical fixed point.  Agreement of theory with experiment implies that the 
critical fixed point studied theoretically is the same fixed point observed in the experiment.  
For the integer quantum Hall transition \cite{klitzing80} the value of the critical exponent $\nu$ that describes 
the divergence of the localization length $\xi$ of the electrons has 
now been measured to an accuracy of a few percent \cite{li05,li09}, so comparison with experiment 
has become a stringent test of theory. One of the most surprising aspects of the  integer quantum Hall transition
is the agreement between experiment and theory \cite{milnikov88, huckestein90b, huckestein95, cain03} 
    concerning the value of the critical exponent. The  agreement is surprising because the Coulomb 
interaction between the electrons is not included in the theoretical models 
\cite{schweitzer84,ando85,chalker88}.  It suggests that the Coulomb interaction is not 
relevant at the integer quantum Hall  critical point.  Yet, this is immediately contradicted 
when we look at the dynamical critical exponent $z$.  For models of non-interacting 
electrons $z$ is exactly 2 \cite{huckestein99}, whereas the measured value is \(\approx1\) \cite{li09}.   

We show in this paper that previous theoretical work has significantly underestimated 
the critical exponent $\nu$ for non-interacting electrons and that there is in fact a 
clear disagreement with experiment.  This suggests that models of non-interacting 
electrons cannot explain the critical phenomena of the integer quantum Hall effect.

Non-interacting electron models of the integer quantum Hall transition are also of 
interest in their own right.  In particular, though the critical field theory for 
such models is as yet unknown \cite{tsvelik07}, it has been speculated that it should 
have conformal invariance \cite{cardy96}.  If so, a scaling relation follows 
\cite{janssen01}, which we test by comparing  recent multi-fractal analyses 
\cite{obuse08,evers08} with our work.

The integer quantum Hall effect \cite{klitzing80} occurs in two dimensional 
electron gases that are subject to a large perpendicular magnetic field.
The application of the field results in the quantization 
of the kinetic energy of the electrons and the formation of Landau levels.
Impurity scattering causes the Landau levels to broaden into Landau bands.
The states at the center of the Landau band are critical while other states are 
localized. 
A quantum Hall transition between quantized values of the Hall resistance occurs whenever the Fermi 
energy passes through the center of a Landau band \cite{aoki81,huckestein95}. The
critical exponent $\nu$  describes the divergence of the localization
length  
\[
\xi\sim |x-x_\mathrm{c}|^{-\nu}\,,
\]
near the critical energy. Here, $x$ can be any control parameter, for example the Fermi energy, that
drives the two dimensional electron gas
though the transition at $x_c$.
In common with other continuous phase transitions, the value of the critical exponent 
is expected to exhibit a high degree of universality \cite{cardy96}.

We have performed a finite size scaling (FSS) analysis of the quantum Hall 
transition in the Chalker-Coddington network model \cite{chalker88, kramer05}. 
In this model the motion of the electron in a random potential and quantizing magnetic field 
is replaced by the transmission of an electron through a network of links and nodes. 
The links describe electron motion along lines of constant potential and 
the nodes describe the scattering of electrons at saddle points of the potential.
The Coulomb interaction between the electrons is ignored.

We use the transfer matrix method to estimate the smallest positive Lyapunov exponent
of a very long network consisting of $L$ layers. Each layer consists of two sub-layers: one a transverse array of $N$ nodes of type A and the other a transverse array of $N$ nodes
of type B.
A transfer matrix $T_{l}$ of size $2N \times 2N$ relates the $N$ right-going and $N$ 
left-going flux amplitudes at the left of the layer to the similar quantities at the 
right of the layer
\[
\left( {\begin{array}{*{20}{c}}
   {{{a'}_1}}  \\
   {{{b'}_1}}  \\
    \vdots   \\
   {{{a'}_N}}  \\
   {{{b'}_N}}  \\
\end{array}} \right) = {T_l}\left( {\begin{array}{*{20}{c}}
   {{{a}_1}}  \\
   {{{b}_1}}  \\
    \vdots   \\
   {{{a}_N}}  \\
   {{{b}_N}}  \\
\end{array}} \right) \,.
\]
The explicit form of the transfer matrix is
\[
{T_l} = B{V_l}A{U_l}.
\]
The scattering at nodes of type A is described by the matrix
\[
A=
\left[
\begin{array}{cccc}
\begin{array}{cc}
1/t & r/t \\
r/t & 1/t
\end{array} & 0_2 &  \cdots & 0_2\\
 0_2 & \begin{array}{cc}
1/t & r/t \\
r/t & 1/t
\end{array} & \ddots& 0_2 \\
 \vdots & \ddots & \ddots & \vdots\\
 0_2 & \cdots & 0_2 &
 \begin{array}{cc}
  1/t & r/t \\
  r/t & 1/t
 \end{array}
\end{array}
\right].
\]
Here, $0_2$ is the $2\times 2$ zero matrix, 
$t$ is the probability amplitude with which current from the left is transmitted 
to the right, 
while $r$ is the amplitude with which current from the left
is reflected back to the left.
Nodes of type B are obtained from nodes of type A by a rotation of 90$^\circ$.
Imposing periodic boundary condition in the transverse direction we then have
\[
B=
\left[
\begin{array}{ccccc}
1/r & \begin{array}{ll} 0\,\, & \,\,0\end{array} & \cdots &  \cdots & t/r \\
\begin{array}{cc} 0 \\ 0\end{array} &
\begin{array}{cc}
1/r & t/r \\
t/r & 1/r
\end{array} & 0_2 &\cdots & \begin{array}{c} 0 \\ 0\end{array}\\
\vdots & 0_2 & \ddots & \ddots & \vdots \\
\begin{array}{c} 0 \\ 0\end{array} &\vdots  & \ddots &
\begin{array}{cc}
1/r & t/r \\
t/r & 1/r
\end{array} & \begin{array}{c} 0 \\ 0\end{array}\\
t/r & \begin{array}{ll} 0\,\, & \,\,0\end{array}
& \cdots &\begin{array}{ll} 0\,\, & \,\,0\end{array} & 1/r
\end{array}
\right].
\]
The transmission and reflection amplitudes are parameterized by the
parameter $x$ according to
\[\begin{array}{*{20}{c}}
   {t = {{\left( {\exp \left( {2x} \right) + 1} \right)}^{ - {1 \mathord{\left/
 {\vphantom {1 2}} \right.
 \kern-\nulldelimiterspace} 2}}}} & , & {r = \sqrt {1 - {t^2}} } . \\
\end{array}\]
The parameter $x$ may be interpreted as the energy of the electron measured
from the Landau band center scaled by the Landau band width.
The matrix $U_l$ is a diagonal matrix with elements
\[
{\left( {{U_l}} \right)_{m,n}} = {\delta _{m,n}}\exp \left( {{\rm{i}}{\varphi _{l,m}}} \right).
\]
and similarly $V_l$ is a diagonal matrix with elements
\[
{\left( {{V_l}} \right)_{m,n}} = {\delta _{m,n}}\exp \left( {{\rm{i}}{{\varphi '}_{l,m}}} \right).
\]
Since the distances between the nodes are random, we suppose that the phases 
are independently and uniformly distributed on $[0,2\pi)$.
(These random numbers were generated using the routine ran described in Chapter B7 of \cite{NRfortran90}.) 

The transfer matrix of the network model is equal to the product of the transfer 
matrices of the layers
\[
T = \prod\limits_{l = 1}^L {{T_l}}.
\]
From this matrix we can define a hermitian matrix $\Omega$ by
\[
\Omega  = \ln \left( {{T^\dag }T} \right).
\]
As a consequence of the conservation of flux the eigenvalues of 
this matrix occur in pairs of opposites sign
\[
\left\{ { + {\nu _1}, \cdots , + {\nu _N}, - {\nu _N}, \cdots , - {\nu _1}} \right\}
\,,\, \nu_1>\cdots >\nu_N >0 \,.
\]
We are interested in the smallest positive Lyapunov exponents which is obtained
in the following limiting procedure
\[
\gamma  = \mathop {\lim }\limits_{L \to \infty } \frac{{{\nu _N}}}{{2L}}.
\]
The value obtained is independent of the 
particular sequence of transfer matrices in the transfer matrix product for 
almost all sequences.
Truncation of the transfer matrix multiplication at a large but finite $L$ yields an
estimate of the Lyapunov exponent with a known precision.
Repeated QR factorizations are needed to avoid a loss of 
precision due to round off error \cite{kramer93}.

To analyze the simulation data we assume that 
the dimensionless quantity
\[
\Gamma  = \gamma N.
\]
obeys an FSS law
\[
\Gamma  = {F_0}\left( {{N^\alpha\left( x-x_{c}\right) }} \right).
\]
Here, $\alpha$ is the reciprocal of the critical exponent $\nu$. In practice, the data deviate from this law. 
This is taken into account by including corrections 
to scaling that arise from non-linearity of the scaling
variables and irrelevant scaling variables. 
In previous work \cite{slevin99a} on the 3D Anderson
model, we have found that the fitting formula
\begin{equation}
\Gamma  = {F_0}\left( {{N^\alpha }u_0 } \right) + {F_1}\left( {{N^\alpha }u_0 } \right){N^y}u_1 ,
\label{fiteq}
\end{equation}
works well. Here,
$u_0 \equiv u_0 \left( x \right)$ and $u_1  \equiv u_1 \left( x \right)$ are, 
respectively, relevant and irrelevant scaling variables,
and $F_0$ and $F_1$ are corresponding scaling functions.
The exponent $y$ associated with the irrelevant variable
is negative, $y<0$.
The relevant variable is zero at the critical point.
For the Chalker-Coddington model the critical point is known
to be exactly $x_c=0$ and
does not need to be found by fitting the numerical data, so 
$u_0 \left( 0 \right) = 0$.

When periodic boundary conditions are imposed, swapping $t$ and $r$,
and translating the network 
so that the A and B nodes are interchanged yields an identical network.
As a result of this symmetry $\Gamma$ is an even function of $x$.
(Note that this holds only for periodic boundary conditions.)
To reflects this, we make $F_0$, $F_1$ and $u_1$ 
even, and $u_0$ odd, functions. 
The condition on $u_0$ follows because the relevant variable is zero at the critical
point.

\begin{figure}[]
\includegraphics[width=\linewidth]{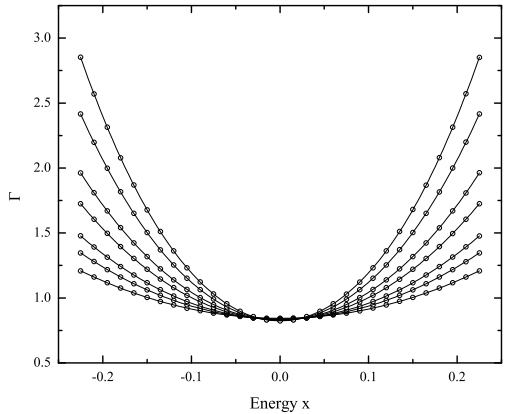}
\caption{Simulation data and FSS fit for the Chalker-Coddington model. 
The different lines correspond to different 
numbers of nodes $N=16, 24, 32, 48, 64, 96, 128$.}
\label{fig1}
\end{figure}

\begin{figure}[]
\includegraphics[width=\linewidth]{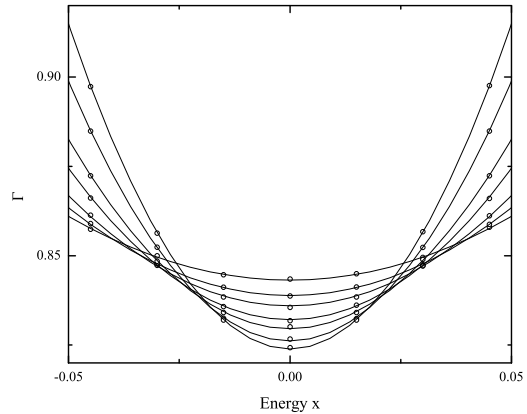}
\caption{The same as Figure 2 but with the focus on data near the critical point so as to make
the existence of corrections to scaling due to an irrelevant scaling variable evident.}
\label{fig2}
\end{figure}

We are able to fit our numerical data to (\ref{fiteq}) provided that data for 
smaller system sizes ($N<16$) are excluded.
All the functions appearing in (\ref{fiteq}) are expanded in Taylor series.
The results of the fit are summarised in Table \ref{table1}.
The data and the fit are plotted in Figs. \ref{fig1} and \ref{fig2}.
The confidence intervals have been obtained from Monte Carlo simulation \cite{NRfortran}.

It is clear from Fig. \ref{fig2} that corrections to scaling are of the order of a
few percent for the smallest $N$.
The precision of our data is $0.03\%$, which requires 
$L\approx 10^8 \sim 10^9$ transfer matrix multiplications to achieve.
Such high precision is required because the dependence on energy of 
$\Gamma$ is very weak near the critical point for the available $N$
and the critical exponent is, in effect, estimated from the variation of the 
curvature with $N$.

We have tested the stability of this fit when the data are
filtered versus the range of $\Gamma$ (Table \ref{table2}), 
the range of $N$ (Table \ref{table3}) and
also checked for stability against increases in the
the order of the various Taylor expansions (Table \ref{table4}).
In all cases there is a large overlap of the confidence intervals with the fit 
in Table \ref{table1}.

\begin{table}
\begin{center}
\begin{tabular}{ccc}
                                                        & estimate      & 95\% confidence interval        \\
$\alpha$                                        & 0.3857        & $\left[0.3849,0.3866\right]$         \\
$\Gamma_{\mathrm c}$            & 0.780         & $\left[0.767,0.788\right]$    \\
$y$                                                     & -0.17         & $\left[-0.21,-0.14\right]$      \\
\end{tabular}
\caption[]{The least squares fit of the FSS model
to the simulation data. The number of data points is $N_{\mathrm D}=217$ and the number of 
parameters is $N_{\mathrm P}=9$. The minimum value of $\chi^2=199.8$ and the goodness of fit $p=0.6$. 
The series for $F_0$, $u_0$,  $F_{1}$ and $u_1$ were truncated at orders 6, 3, 0 and 2, respectively.}
\label{table1}
\end{center}
\end{table}

\begin{table}
\begin{center}
\begin{tabular}{ccccc}
$\Gamma<2.0$ & 0.3858 &  $\left[0.3850,0.3867\right]$ & 0.779 & $\left[ 0.767, 0.788\right]$ \\
$\Gamma<1.5$ & 0.3855 &  $\left[0.3847,0.3864\right]$ & 0.781 & $\left[ 0.769, 0.790\right]$ \\
$\Gamma<1.0$ & 0.3826 &  $\left[0.3796,0.3858\right]$ & 0.789 & $\left[ 0.778, 0.797\right]$ \\
\end{tabular}
\caption[]{The best fit of $\alpha$ and $\Gamma_c$ with 95\% confidence intervals.
The data have been filtered using the condition on $\Gamma$ at the left.}
\label{table2}
\end{center}
\end{table}

\begin{table}
\begin{center}
\begin{tabular}{ccccc}
$N\ge 24$ & 0.3854 & $\left[0.3845,0.3864\right]$ & 0.782 & $\left[0.763,0.794\right]$ \\
$N\ge 32$ & 0.3849 & $\left[0.3837,0.3861\right]$ & 0.787 & $\left[0.768,0.800\right]$\\
\end{tabular}
\caption[]{The best fit of $\alpha$ and $\Gamma_c$ with 95\% confidence intervals.
The data have been filtered using the condition on $N$ at the left.}
\label{table3}
\end{center}
\end{table}

\begin{table}
\begin{center}
\begin{tabular}{ccccc}
$F_0$ & 0.3858 & $\left[0.3849,0.3866\right]$ & 0.779 & $\left[0.767,0.788\right]$ \\
$u_0$ & 0.3853 & $\left[0.3843,0.3864\right]$ & 0.780 & $\left[0.767,0.789\right]$ \\
$F_1$ & 0.3828 & $\left[0.3754,0.3917 \right]$ & 0.780 & $\left[0.767,0.789 \right]$\\
$u_1$ & 0.3857 & $\left[0.3849,0.3865\right]$ & 0.780 & $\left[0.767,0.788\right]$\\
\end{tabular}
\caption[]{The best fit of $\alpha$ and $\Gamma_c$ with 95\% confidence intervals.
The order of the expansion of the indicated function
has been increased by two compared with TABLE \ref{table1}.}
\label{table4}
\end{center}
\end{table}
Our FSS analysis yields the following best fit value and $95\%$ confidence interval for the critical exponent
\begin{equation}
\begin{array}{{c}{c}}
   {\nu  = 2.593} & {\left[ {2.587,2.598} \right]}  .\\
\end{array}
\label{nuestimate}
\end{equation}
This result is consistent with but also a considerable improvement on the original estimate  $\nu=2.5\pm.5$ \cite{chalker88} 
of Chalker and Coddington.

Our estimate is significantly larger than the oft quoted result of   $\nu = 2.34\pm .04$ of Huckestein and Kramer \cite{huckestein90b} for a
random Landau matrix model.
This disagreement could be taken as evidence against the universality of the critical exponent. However, we feel that a more likely explanation is that the precision claimed by Huckestein and Kramer 
   is too optimistic. This may be clarified in future work.

Our result is also different from the analytical result $\nu= 7/3$ of  Milnikov and Sokolov \cite{milnikov88}. However, this value is not expected to describe the true  critical point but an intermediate behavior  not too close to the critical point.

Our result also  disagrees with   \(\nu = 2.37\pm0.02 \) of Cain \textit{et al.} \cite{cain03} based on a real space renormalization group approach \cite{galstyan97}. A more recent calculation \cite{mkhitaryan09} on a network with a triangular lattice gave \(\nu\approx2.3\sim2.76\). Thus, the precision claimed by Cain \textit{et al.} may be questionable. Moreover, this approach involves an uncontrolled approximation and so the discrepancy may not be significant.

More important than the discrepancy with previous theory is the disagreement 
of our estimate with recent experiments.
Li {\it et al} \cite{li05, li09} 
measured  $\nu=2.38\pm.06$ in experiments on GaAs-AlGaAs 
heterostructures. In these experiments, an exponent \(\kappa=1/\nu z\) was measured. Extraction of the critical exponent \(\nu\) requires an independent measurement of the dynamical exponent \(z\). Li {\it et al} measured \(z\approx1,    \) but seem to have obtained their result for \(\nu\) by supposing \(z=1\). An unambiguous comparison with theory will require a more careful consideration of the precision of the estimate of \(z\).

 In our opinion, the coincidence of previous theoretical and experimental estimates 
of \(\nu \) is not significant. Non-interacting theory predicts \(z=2\) in clear disagreement with experiment.  Calculations \cite{huckestein99} 
within the Hartree-Fock approximation suggest that the observed value of $z\approx1$  
may be explained by including the  Coulomb interaction.
 We speculate that the discrepancy in the value of $\nu$ that we report here may also be explained in this way.  

Assuming that the  critical field theory of the integer quantum Hall effect has  conformal invariance,  
it has been shown that $\Gamma_c$ and $\alpha_0$ (determined in multi-fractal 
analysis \cite{obuse08,evers08}) are related by \cite{janssen01}
\begin{equation}
{\Gamma _c} = \pi \left( {{\alpha _0} - 2} \right).
\label{conformal}
\end{equation}
Eq. (\ref{conformal}) follows from a conformal mapping
between the strip used in transfer matrix calculations and the 2D plane used
in multi-fractal analysis. 
Inserting our result for $\Gamma_c$ into this formula we obtain
\[
\begin{array}{cc}
   \alpha_0 = 2.248  & [2.244,2.251] \\
\end{array}.
\]
This value is not consistent with either
$2.2617 \pm 0.0006$ reported in \cite{obuse08} or $2.2596 \pm 0.0004$ in \cite{evers08}.
However, we cannot conclude that the critical theory does not have conformal invariance because the leading correction to scaling decays very slowly with $N$, as is
indicated by the small value of the irrelevant exponent $y$.
This does not affect the estimation of  the critical exponent $\nu$ but it does complicate the estimation of $\Gamma_c$ and, probably, $\alpha_0$.

In conclusion, we report an estimate of the critical exponent \(\nu\) of the integer quantum Hall effect that is significantly larger than both previous theoretical estimates and, more importantly,  experimentally measured values. We speculate that models of non-interacting electrons cannot explain the 
critical phenomena of the integer quantum Hall effect.  
Further work is needed to come to a definite conclusion concerning the conformal 
invariance of the critical theory of non-interacting electron models of the 
integer quantum Hall transition.

This work was supported by Grant-in-Aid No. 18540382.
We would like to thank H. Obuse, A. Furusaki, F. Evers and A. M. Tsvelik
for helpful comments.

\bibliography{references}

\end{document}